# Bucking the Trend: An Agentive Perspective of Managerial Influence on Blog's Attractiveness

Carlos Denner dos SANTOS
University of Brasília
carlosdenner@unb.br

Isadora V. Castro
University of Brasilia
isadora.castro@gmail.com

George Kuk
Nottingham Trent
george.kuk@ntu.ac.uk

Silvia Onoyama
EMBRAPA/Brazil
silvia.onoyama@embrapa.br

Marina Moreira
University of Brasilia
marinamoreira@unb.br

## Abstract

*Blog management is central to the digitalization of work. However, existing theories tend to focus on environmental influence rather than managerial control of a blog's attractiveness at a microlevel. This study provides an agentive account of the adaptive behaviours exerted by the bloggers through the ways they use contents of their blogs to locate and harness their structural network positions of a blogosphere. We collated individual characteristics of 165 bloggers who blogged about economics, and then analysed the ways they maintained the contents of their blogs. We used network analysis and monomial logistic regression to test our model predictions. Our findings show that in contrast to less attractive blogs, bloggers who are mindful of their peers' contents as a means of maintaining network positions attract a significantly higher level of traffic to their blogs. This agentive perspective offers practical insights into how nodal preferences can be reversed in blog management. We conclude the paper by discussing contributions to theory and future research.*

## 1. Introduction

Several studies have discussed blog management, structure and environment. In addition to the structural capital perspective [1], existing literature seems to prescribe an environmental perspective as a main mechanism of how blog management is being externally governed [2]. This mechanism means that the attractiveness of blogs tends to concentrate on a few and is irreversible [3].

While attractive blogs are favored indefinitely, the majority is prone to failure, overlooking actions taken by bloggers [4]–[6]. However, the everyday practice of blog management seems to suggest otherwise. In an attempt to reconcile theory and reality, we ask: how the characteristic and managerial actions taken by a blogger influence a blog's attractiveness? To do so, in this study, we combined the network and organizational perspectives to explain the variation of blog's success by taking into account the managerial actions taken by the bloggers [7]–[10].

Although blog management may be restricted by contexts, the effects of the environment offer a partial explanation. An agentive perspective based on bloggers' actions often runs counter to the complexity and dynamic characteristics of the organizations [10], [11]. Despite prior research on blogs [4], [5], [11]–[13], how blog management at a micro-managerial level plays out at a network level is less understood.

It is relevant to clarify these relationships, not only within the framework of social media, but, especially, in organizational theories. When we understand which mechanisms are responsible for the evolution of an organization, we can ascertain how relationships are established and developed [14]. We examine the influence of managerial actions exerted by bloggers against environmental contingencies on the performance of these organizations [11], [15].

There are many theoretical perspectives that seek to explain the relationship between managers, organizations and environment [7]. From an organizational theory point of view, our research question draws upon a frequent debate between selection and adaptation from Organizational Ecology and Structural Contingency. We used social network analysis (SNA) to analytically test the links between managerial action and network position in terms of structural capital [1]–[3].

## 2. Blogs: management and the network

Irrespective of its functions, a blog can be designed for entertainment or served as a business tool [16]. The number of its subscribers and visitors (audience) represent indicators that can demonstrate its capacity to produce material of interest. This is why such indicators are used by bloggers and related companies as one of the main ways of evaluating blog performance or attractiveness [17], [18].

In addition to managing their own pages, bloggers also read other blogs and, based on this, organize and make publicly available lists of recommended links that they consider relevant. These recommendations, describe different network pages, known as Blogosphere, making it possible to visualize the connections [19]. This network functions like a society of blogs and is based on following agreed-upon practices to ensure co-existence of an online community [20].

Structurally-speaking, a connection preference mechanism is the main reason for forming these relationships [6], [13]. More attractive blogs tend to become increasingly more attractive [2], [12], [21], generating a centralized structure, that is divided between central blogs, which hold a greater number of relationships whereas peripheral blogs hold fewer established relationships.

## 3. Management and network theories

According to Organizational Ecology, organizations are at the mercy of the environment, in that failure or success is explained not in terms of managerial actions, but selective environmental processes [9]. The role of the managerial action is limited by four primary factors: (1) the organizational structure, which constrains individual behavior; (2) the lack of resources, which limits management activities; (3) the competition for resources, which reduces the possibility of strategic choices; and (4) the supremacy of the environment above individual rationality based on the assumption that the process of natural selection operates in the "collective". In this respect, although prior research [9] does not dismiss managerial influence, the role of a manager is very often seen as being inactive or symbolic [7].

From the perspective of structural contingency, despite the indirect effect of the environment, organizational changes that are initially caused by internal factors will require internal structures to be realigned within the environmental contexts [10], [22]. Variations in the internal structures are found to be contingent with the environment. Often managerial actions are required to meet the environmental demands by functionally realigning internal structure to an external context [10], [23].

However, at the network level, an organizational structure captures the relationships between actors [21]. This perspective takes an agentive account based on actions and interactions between its network participants, which forms the basis of structural capital based on a micro-level of analysis, denoting the benefits of different network positions [21].

From an environmental delineation point of view, networks comprise an open and dynamic system, where new actors are added and excluded all the time, and where their performance is not only influenced, but also determined by a connection preference mechanism [2], [4], [12], [21], [24]. It produces an effect known as "the rich get richer" [2], that imposes itself in a deterministic fashion [9], [25], whereby the actors are divided into two groups: (i) few individuals that are attractive, represent high levels of performance and tend to attract more connections over time; and (ii) the many individuals that are not attractive, result in low levels of performance and few connections. It presupposes new entrants connect to those more connected, so a novice blogger has to do is to wait for successful blogs to be abandoned, then set this up as a passive agent to establish the performance of his or her page, since the survival or death of a blog [9], [25] occurs irrespective of its adaptation capacity [26].

On the other hand, research into structural capital notes how a network structure and, consequently, the position of actors in this network, can produce performance variations, involving access to resources, time availability and access to information. One of the features that allows this variability is the actors' ability to visualize existing structure of the network [21]. In the case of blogs, it is possible to observe this structure on recommendation lists available [19], [21].

Besides being viable, this analysis is highly interesting, since the network structure is public, so understanding and exploring prominent positions can be achieved, not only by media managers, but also by interested companies acting as partners [16], [21]. This statement supports the structural capital analysis of each actor for calculations to be made in terms of the centrality of degree, proximity, intermediation and PageRank [27], [28], as shown in Table 1.

Depending on the measure utilized, centrality can be construed from different viewpoints [21], [27]. Because of their ability to predict the performance of a network participant, these measures also inter-relate in mathematical terms [31]. However, it is important to stress that different measures are not the same, not even in mathematical terms, nor in theoretical terms, despite being correlated [1], [21], [24], [32].

### 3.1. Modelling blog attractiveness

It is possible to observe the outcome of blog's adjustment capacity and context (org. performance) by the size of its readership [18], [20], [28].

Table 1. Centrality measures in social networks

| Structure | Definition | Definition | Source |
|---|---|---|---|
| Centrality of Degree (Degree) | Number of direct connections established with other network actors | $\sigma_D(x) = \sum_{i=1}^{n} a_{ix}$ | [27] |
| Centrality of Closeness (Closeness) | Average number of steps required to access all other network actors | $\sigma_C(x) = \dfrac{1}{\sum_{i=1}^{n} d_G(x,i)}$ | [27] |
| Centrality of Intermediation (Betweenness) | Average number of times an actor acts as a bridge between other network actors | $\sigma_B(x) = \sum_{i=1, i \neq x}^{n} \sum_{j=1, j<i, j \neq x}^{n} \dfrac{g_{ij}(x)}{g_{ij}}$ | [27] |
| PageRank Algorithm | Total number of recommendations according to the popularity of those making the recommendations | $PR(i) = \sum_{j \to i} c \, \dfrac{1}{d_j} \, PR(j) + (1-c)$ | [30] |

Source: Developed by the authors.

In a sense, a blog's attractiveness is influenced by the structural capital [21], [27], which is instigated by the blogger who acts and explores his structural capital in the network to seek gains [21]. Regarding the attractiveness of structural capital on blogs, the number of recommendations received (or its input centrality), represents traction in terms of greater visibility because recommendations are public, and bestow the blog with legitimacy [20], [33]. A greater degree of centrality input increases access to resources through its vast network connections [27], such that occupying a central network position positively influences blog attractiveness.

- (H1.1) a higher degree of input centrality positively and significantly influences attractiveness.

Since the degree of closeness is also a synonym of speed of access [27], blogs that are a few steps away from others access network information faster. These blogs quickly become aware of new innovations and partnership opportunities to be explored, and are also, inversely, capable of reaching their audiences much faster when they are the first to promote content [12].

- (H1.2): a greater degree of proximity positively and significantly influences attractiveness.

Blogs with a high degree of intermediation serve as bridges between different communities. The importance of a position of intermediation is reinforced when it operates as a network resilience measure, that is to say, how important is the survival of a blog with a high degree of intermediation to maintaining network interconnectivity [24]. Thus, this position, in addition to reinforcing the role of the actor as a hub of communication with other actors [12], [34], positively influences the attractiveness, since this position confers advantages on a blog, not only in terms of what it can publish, but also to control what information other pages can access.

- (H1.3): a higher degree of intermediation positively and significantly influences attractiveness.

PageRank denotes recognition, not only by other blogs, but also considers recommendations made by other types of pages [14], [32]. As the measure varies between 0 and 10 [30], blogs with values nearer 10 are the most cited on the Internet and are those that present the most popular contents [14], having a higher probability of attracting readers [35]. In this sense, its value positively influences the attractiveness of a blog, since the relevance indicated by the algorithm can be understood to be a form of legitimization.

- (H1.4): a higher value of PageRank positively and significantly influences attractiveness.

### 3.2. Characteristics and overall actions

Unlike classic organizations, blogs are peculiar in the sense that, even though there is a high turnover of co-authors, the person responsible for establishing, outlining and reinforcing the objectives of an organization over time [8] is almost always a permanent fixture [36]. Even though different collaborators may no longer take an active part in drafting the page, in most cases the idealizer and manager of the blog is the main person responsible for the financial, administrative and strategic administration of the page, taking care of different aspects, from the formation of business partnerships to keeping up hospitality payments and designing the layout [12], [20], [36].

In this sense, a greater period of work experience enables a blogger, not only to understand the workings and good practices of the community to which he belongs, but also to learn lessons of success and failure by studying the past history of his own blog and the development of other pages, as well as to acquire knowledge and improve his or her skills [37]. Managerial experience can also maximize the benefits of having a prominent position within the network regarding audience, since an experienced blogger can act according to the lessons learned, thereby mitigating the effects of environmental vulnerabilities.

- (H2): a longer work experience positively influences the relationship between structural capital and attractiveness.

Present literature also points to the fact that bloggers who have a professional activity related to the area of blogs, are agents who can make decisions about the subject in a more critical and reliable way [15], [38]. In addition, a blog visitor will notice in the quality of arguments [39], since good practice of blog management can influence the arguments used in the texts. In order for a blogger to maximize advantages over competitors, when dealing with the inflow of readers, it is important that a blogger uses his professional status publicly to signal quality [38].

Blogs differ from one to another and it is up to the reader to decide which page to use to find the most reliable information [14]. Thus, a professional performance will enable the blogger to take advantage of the effects of structural capital as, for example, in the positive impact created by having a greater number of recommendations in the number of accesses received by the blog.

- (H3): professional performance in the area positively moderates the relationship of structural capital and attractiveness.

A blog only exists when there are postings, since, like social media, the dissemination of content is the first act of its creation [5]. In this sense, irrespective of its frequency or quality, content publishing is indispensable for a blog to be seen as active. However, in the search not only for survival, but also for success, bloggers can exercise this function strategically. Also frequent content publishing can inspire confidence among its readers [20]. Against this, a higher content publishing tends to be seen positively, making it possible to take advantage of the benefits of strutural capital by attracting more readers.

- (H4): greater content publishing positively influences the relationship between structural capital and attractiveness.

Audience communication is a primordial function. Interviews carried out with Brazilian bloggers reveal that knowing the opinion of their readers about what is being published is one of the main reasons for the continuity of a page [18]. This feedback makes it possible for the blog readers, not only to receive content passively, but also to help them to see this as being valid, legitimate, trustworthy and pertinent [40]. It is through these responses to feedback that bloggers establish interaction and communicate with their readers [41]. In content production, the presence of postings is essential. However, interaction by means of feedback, in spite of being a desirable element, is arbitrary [18], [42]. But when a blog allows comments to be written, an expectation for interaction is created. In this situation, a blogger has two options: either (a) to meet these expectations and establish communication, either by responding to a query or exchanging information; or (b) not to respond to the comment [41]. The first option is clearly beneficial, since, in addition to fulfilling his purpose, the blogger can explore the comments space to suggest other publications on the blog, explain a point of view in a better way, and to learn more about the needs and preferences of his or her readers [43].

On the other hand, making a communication channel available, but ignoring the blog reader, can generate, not only dissatisfaction, but also rejection [41]. An isolated action does not imply severe consequences, but, if failing to communicate becomes habitual to the point that a blog reader feels jeopardized, this can lead to a loss of readers and, consequently, to organizational failure. Thus, audience communication, instead of merely avoiding negative effects, involves gains, enabling a blogger to mitigate adverse underprivileged positions.

- (H5): audience communication positively influences the relationship between structural capital and attractiveness.

The characteristics and managerial actions of bloggers are capable of actively and positively influencing the impact that structural capital exerts over the blog attractiveness. The strategic use of managerial elements can increase the advantages gained by occupying central positions that allow for better visibility and access to resources (Figure 1).

**Figure 1. Blog attractiveness**

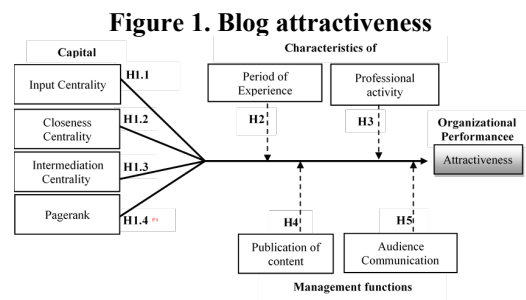

Source: Developed by the authors.

## 4. Methods

This research method used various sources of secondary data and involved three phases. In the first phase, the empirical context and sampling were established. For this, a network such as one with a "snowball" effect, was outlined: firstly, a blog was selected, based on its worldwide representativeness, among Economy blogs; then, a priori, the network boundary was established as being at 1 degree of separation, which meant considering a network formed by the blog chosen initially, all the blogs cited by it, as well as all the relationships between them [33]. In all, a sampling of 165 Economy blogs was obtained, which are responsible for disseminating information and encouraging debate about issues such as finance, inflation and unemployment. The sampling was not probabilistic and followed the criteria of accessibility.

In the second phase, a collection of secondary data was carried out. In the case of attractiveness, the number of visits attracted during a period of six months, between July and December 2014, was used as a proxy, by using the Similar Web website, specialized in mapping visitor traffic. The heterogenic behavior of the attractiveness variable was checked, since there were some blogs that showed 500 accesses and others with nearly 1.6 million visits during the same time interval. However, for the analysis, the variable was classified into five different categories: very low (initial 20% of classified data), low (20% to 40%), average (40% to 60%), moderately high (60% to 80%) and high (final 20% of classified data). We then referred to the value of PageRank, also through a website called PageRank Calculation, which consists of a virtual algorithm calculator, so that 165 blogs were awarded values between 0 and 10 for the referred variable. The period of experience and professional activity variables were collected by means of investigations into the public profile of the bloggers, so that the first was calculated based on years and the second, as the predominate profession declared. A set of dummies was then created to indicate each one of the professional activity categories: (1) teacher, (2) economist, (3) businessman/CEO, (4) journalist, (5) consultant/investor, and (6) others. With regards to managerial functions, the content publication was measured by means of the number of publications made during a period of seven days, which can indicate how frequently the content is produced. We then observed if the posts published during this seven-day period contained any feedback. Thus, in relation to audience communication, the value of 1 was conceived for when at least one reply was sent to the readers, and 0 for when this did not occur.

In the third phase, a data analysis was carried out. A social network analysis measured the variables related to structural capital, while Ucinet software (V6) was used to calculate values relative to the independent variables, input centrality, closeness and intermediation [45], considering all the combinations of 165 blogs. The values extracted were integrated into the data base. We then carried out a descriptive analysis and, finally, a multinomial logistic regression to test the hypotheses pictured in fig. 1.

The choice of a multinomial logistic regression was justified by the fact that the dependent variable is nominal and presents more than two categories [46], [47]. This technique makes it possible to calculate the conditional probability of occurrence of each category of a dependent variable, based on a combination of independent variables [48], in accordance with the following equation, where Y= attractiveness; j=categories; wi= structural capital and manager characteristics variables:

$$\text{Prob}(Y_i = j | w_i) = \frac{\exp(w_i' \alpha_j)}{\sum_{j=1}^{J} \exp(w_i' \alpha_j)}$$

## 5. Results

The blogs analyzed presented heterogeneity in terms of attractiveness. Those considered to have a low level of attractiveness obtained between 500 and 6.800 visits; those classified as average had between 6.800 and 18 thousand visits; the moderately high, from 35 to 102 thousand; and, finally, those ranked high, between 102 thousand and up to 1.6 million visits. Regarding input centrality, a blog receives, on average, 6 recommendations from others. Regarding closeness centrality, blogs are, on average, five steps away from all others on the network.

With regards to centrality of intermediation, the values found vary between 0 and 8.285, with an average of 160.22 and a standard deviation of 705.85. The value of PageRank, in spite of having an average of approximately 5, showed a variance in the sense that 87.8%, or 145 of the 165 blogs had a value equal to or below 6. The research involved bloggers who had up to 14 years of experience. Even though we found inexperienced bloggers, who had less than one year of experience, most have, on average, 6 years of experience. The heterogenic character of the network can be seen even by the variations related to content publication. During a period of seven days, blogs were found that had not published anything, and pages that had up to 84 postings. It was found that 70.3% 116 blogs, did not carry out audience communication through feedback and comments. As for professional activity, bloggers of 141 pages carry out an activity in the subject area of the blog. 85.5% of bloggers investigated had basic knowledge about Economics, obtained formally or informally, by lecturing in a faculty of economics or had learned it through personal experience. On the other hand, 14% only operate as bloggers within the sphere.

The higher this is, greater are its chances of being classified between moderately high and high, not rejecting hypotheses 1.1. In that which refers to closeness centrality, for each additional point of the variable, greater is the chance that these will be classified as low attractiveness (RRR 1,75) and average (RRR 2,15) in relation to very low attractiveness. That is to say, the further away it is, the greater chance there is to have a low audience rating. On the other hand, the closer it is, greater is the chance that this will be classified as moderately high, given the negative value and significance of the coefficient β=-1.21, not rejecting H 1.2.

For centrality of intermediation, this presented a significant relation with low attractiveness (RRR 1,02), it not being possible to conclude that a higher level of intermediation would positively and significantly influence, attractiveness. Hypothesis 1.3 was discarded. Finally, the PageRank value influenced significantly attractiveness, given the positive values of the RRR in moderately high and high attractiveness (5.43/1.30), not rejecting H1.4.

Regarding the effects of managerial intervention, we then carried out a moderation analysis to explain the role of managerial intervention in the 165 blogs (Table 3). As a result of the moderation test for period of experience, statistically significant effects were obtained in relation to closeness centrality and to PageRank attractiveness. For the closeness centrality and attractiveness relationship, each additional moderation point increases the chance of being classified as low attractiveness by 91% (RRR 1.91), and by 26% (RRR 1.26) for average attractiveness, and by 36% (RRR 1.36) for moderately high and by 18% (RRR 1.18) for high attractiveness, in relation to very low attractiveness. The period of experience effect was shown to be significant and positive in relation to the influence that closeness centrality exerts over attractiveness.

Regarding the influence that PageRank exerts over attractiveness, the period of experience showed that it was less likely to be associated with low attractiveness (RRR=0.92) and with moderately high attractiveness (RRR=0.89), in comparison with very low attractiveness. The relationship between PageRank and attractiveness becomes weaker on a longer period of experience. For input and intermediation centrality, the moderation effect on the period of experience was insignificant. Hypotheses 2 is not rejected for the closeness and attractiveness centrality relationship.

In the case of professional activity, significant effects were found for the categories of teacher, economist and consultant. Teacher had a positive effect in the relationship between input centrality and low attractiveness (RRR=1,85) and average attractiveness (RRR=1.31) in comparison with very low attractiveness. It may be deduced that the moderation effect was limited up to a certain level of attractiveness. With regards to the activity of a teacher between closeness centrality and attractiveness, the moderation effect is probably associated with moderately high attractiveness (RRR= 5.16) and less associated with average attractiveness (RRR= 0.41). With regards to the influence that PageRank exerts on attractiveness, the category of teacher is less associated with low attractiveness (RRR=0.34), moderately high attractiveness (RRR=0.13) and high attractiveness (RRR=0.76), in comparison with very low attractiveness, and is associated more with average attractiveness (RRR=1.34). It may be inferred the moderation effect for the teacher was negative.

The category of economist had a positive and significant effect on the relationship of input centrality and low attractiveness (RRR=2.96), and average attractiveness (RRR=1.17). However, it is less likely to be associated with moderately high attractiveness (RRR=0.69). The effect of closeness centrality was increased by the category of economist in cases of low attractiveness (RRR=2.79) and reduced in moderately high attractiveness (RRR=0.84) and high attractiveness (RRR=0.81). The centrality of intermediation and attractiveness relationship was little influenced by the economist category. However, the effect of PageRank on attractiveness was reduced significantly and drastically in all level of attractiveness, since all RRR values for all categories were less than 1.

The activity of consultant/investor is associated more with the relationship between input centrality and low attractiveness (RRR=136), and average attractiveness (RRR=1.97) and less likely to be associated with moderately high attractiveness (RRR=0.26) and high attractiveness (RRR=0.14), when compared to very low attractiveness. With regards to how closeness centrality acts with attractiveness, the category of consultant/investor is less likely to be associated with low attractiveness (RRR=0.68) and average attractiveness (RRR=0.28) and associated more with moderately high attractiveness (RRR=1.73) and high attractiveness (RRR=1.55) in comparison to very low attractiveness. The activity of a consultant/investor, like that of an economist, had a very slight influence on the relationship between centrality of intermediation and attractiveness, since these obtained RRR ratings a little higher than 1. The effect of PageRank decreased with the consultant/investor activity in low levels of attractiveness (RRR=0.52), moderately high levels (RRR=0.11) and high levels (RRR=0.02). Due to the predominance of negative effects in these three categories, H3 was discarded.

In relation to content publication, it was seen that this exerts a significant positive moderation effect on how strongly closeness centrality is more associated to moderately high attractiveness (RRR=3.09) and high attractiveness (RRR=3.26) in comparison to very low. In the case of input and intermediation centrality and low attractiveness, effects were slightly negative and there were no significant effects for PageRank and attractiveness. H4 was not rejected for the relationship between closeness centrality and attractiveness.

The audience communication managerial action is less associated in the relationship between input centrality and a low level of attractiveness (RRR=0.54) and average attractiveness (RRR=0.74). Moreover, these have a negative effect on the relationship between closeness centrality and attractiveness. On the other hand, the effect is positive for relationships between centrality of intermediation and high attractiveness (RRR=1.03) and PageRank and low attractiveness (RRR=1.38), average (RRR=2.15), moderately high (RRR=1.60) and high (RRR=2.63), comparing to low attractiveness. H5 was not reject as centrality of intermediation and PageRank and attractiveness are related. Tables 2 summarizes results.

## 6. Discussions: implications of findings

Research that focuses on structural capital suggests that the centrality of degree (H1.1), of closeness (H1.2), of intermediation (H1.3) and the value of PageRank (H1.4) significantly influence blogs' attractiveness. However, the findings showed that the independent effects of structural capital based on blog attractiveness are influenced by input centrality, by closeness and PageRank, and that the centrality influence on intermediation is irrelevant.

This might be due to the fact that the only explicit relationships between blogs are the lists of recommendations [19]. If a greater number of recommendations in other blogs, that is to say, input centrality, indicate greater visibility and signify legitimacy [20], and if this is a more simple choice of criterion for a reader to decide where to concentrate his or her attention [34], it is reasonable to consider that the privileged position influences performance.

After analyzing the independent effects, a moderation analysis was made of management, the structure and environment of the blogs, which are better explained if considered together. In the first place, the effects of a longer period of managerial experience were examined (H2) as well as professional performance (H3). The results showed a negative intervention for professional performance as regards how strongly structural capital influences attractiveness. As a form of social media that was primarily created as a hobby for bloggers [16], unlike a newspaper, it is possible to consider that the audience takes greater interest in other aspects, such as the importance they give to that text or to the quality and relevance of photographic images used, rather than to whether or not the blogger has a diploma on the subject or even experience. For the blogger, this means anonymity may be preserved without harming credibility in any significant way, which clearly goes against the thinking of Chesney and Su [38], who defend him/a blogger as being of paramount importance.

In contrast, the period of experience presented different behavioral characteristics. With regards to centrality of closeness in relation to attractiveness, it was significantly amplified, showing how important managerial experience is for the strategic administration of resources that can be accessed more rapidly [27]. On the other hand, the longer the blogger's period of experience, the weaker the impact that a high PageRank value can have on reader attraction. If, on its own, PageRank positively influences the number of visits and, if the inclusion of a managerial figure inverts this relationship, then we have evidence to believe that there are levels of experience that are harmful to organizations.

Moreover, if the period of experience was understood, not as a personal characteristic, but as a reflection of the time a blog has existed or its length of service, the outcome would be the existence of connection preference itself. If we strictly follow the views of Barabási [50], a blog is independent of a blogger's efforts [3]. However, it is also purely ficticious to think that it is viable for an organization to exist on its own [10]; to go along with the idea that a blog accumulates resources on its own irrespective of the role of a manager seems innocent. Thus, it is inferred that length of service gives the blog the means to obtain attractiveness because it enables a blogger to understand how to operate in the network, accumulating knowledge and social capital over time.

Having said that, if at the outset, positive managerial influence may be understood as a connection preference synonym, the inversion of such influence that occurs when a blog reaches a certain level of service indicates its relativization. Once a blogger has learnt the best way to operate, the same length of service that initially enables an organization to overcome vulnerabilities and achieve success, later tends to lead to failure or obsolescence, breaking the cycle in which rich actors get richer. Even though this is a recent and little explored issue, especially in the context of blogs [3], exploratively speaking, this may be one of the phenomenon responsible for the change in leadership among attractive blogs over time. 

Furthermore, there are perhaps other correlations that should be further investigated. For instance, one could analyze how the quantity and quality of published content directly influences the blogs' attractiveness. The same applies to experience and other managerial characteristics. Another possibility could be a positive correlation of these variables with the existing structural capital (more experience and more content leads to larger networks). Also, structural capital itself, like PageRank, could be an aspect of the blog's success, hence being a rich terrain for future research as dependent variable.

The effects of managerial functions were tested. Following the recommendations put forward by Barnard [8], that it is up to managers, not only to outline their objectives, but also to use their best efforts to maintain survival and communication, we suggest that the content publication (H4) and audience communication (H5) are aspects capable of influencing the relationship between structural capital and attractiveness. With regards to content publication, findings showed that the number of postings published by a blogger is relevant, and capable of reinforcing a positive relationship between access speed and controlling resources in accordance with the number of blog readers. Even though a blog needs postings to be considered active on the network, the strategic use of publications to attract audiences is optional/discretionary [5]. In the case of audience communication, this was found to be significantly influential in relation to the positive effects resulting from the quantity of recommendations received by the blog, and its attractiveness.

Thus, if a highly recommended blog has a greater chance of attracting readers, then one that is well recommended, combined with a manager who knows how to communicate through feedback, has an even better chance of success. Communicating with audience via feedback, not only avoids negative effects, such as the alienation of readers [41], but also promotes better org. outcomes.

**Table 2. Multinomial Analysis Results**

| Variables | Attractiveness | | | | | | | |
| (n=165) | Low | | Medium | | Moderate | | High | |
| | β | RRR | β | RRR | β | RRR | β | RRR |
|---|---|---|---|---|---|---|---|---|
| Interceptor | 0,19(***) | 1,20 | -0,08(*) | 0,93 | 0,77(***) | 2,16 | 0,84(***) | 2,33 |
| CE | -0,36(***) | 0,70 | 0,00 | 1,00 | 0,53(***) | 1,70 | 0,42(***) | 1,53 |
| CP | 0,56(***) | 1,75 | 0,76(***) | 2,15 | -1,21(***) | 0,30 | -0,19 | 0,83 |
| CI | 0,01(**) | 1,02 | 0,01 | 1,01 | 0,00 | 1,00 | 0,00 | 1,00 |
| PR | 0,43(***) | 1,53 | -0,29(*) | 0,74 | 1,69(***) | 5,43 | 0,26(**) | 1,30 |
| TExCE | 0,03 | 1,03 | -0,03 | 0,97 | 0,01 | 1,01 | 0,01 | 1,01 |
| TExCP | 0,65(***) | 1,91 | 0,23(***) | 1,26 | 0,31(***) | 1,36 | 0,17(***) | 1,18 |
| TExCI | 0,00 | 1,00 | 0,00 | 1,00 | 0,00 | 1,00 | 0,00 | 1,00 |
| TExPR | -0,08(*) | 0,92 | 0,01 | 1,01 | -0,11(*) | 0,89 | -0,05 | 0,95 |
| AP1xCE | 0,62(***) | 1,85 | 0,27(*) | 1,31 | -0,19 | 0,82 | -0,04 | 0,96 |
| AP1xCP | 0,05 | 1,05 | -0,88(***) | 0,41 | 1,64(***) | 5,16 | -0,03 | 0,97 |
| AP1xCI | -0,02(***) | 0,98 | -0,01 | 0,99 | 0,00 | 1,00 | 0,00 | 1,00 |
| AP1xPR | -1,09(***) | 0,34 | 0,30(***) | 1,34 | -2,03(***) | 0,13 | -0,27(***) | 0,76 |
| AP2xCE | 1,09(***) | 2,96 | 0,16(*) | 1,17 | -0,37(***) | 0,69 | -0,12 | 0,89 |
| AP2xCP | 1,03(***) | 2,79 | 0,04 | 1,04 | -0,18(***) | 0,84 | -0,21(***) | 0,81 |
| AP2xCI | -0,02(**) | 0,98 | 0,00 | 1,00 | 0,01 | 1,01 | 0,00 | 1,00 |
| AP2xPR | -2,37(***) | 0,09 | -0,95(***) | 0,39 | -3,46(***) | 0,03 | -0,96(***) | 0,38 |
| AP3xCE | 0,31(***) | 1,36 | 0,68(***) | 1,97 | -1,36(***) | 0,26 | -1,95(***) | 0,14 |
| AP3xCP | -0,38(***) | 0,68 | -1,28(***) | 0,28 | 0,55(***) | 1,73 | 0,44(***) | 1,55 |
| AP3xCI | -0,02(*) | 0,98 | 0,01 | 1,01 | 0,02(**) | 1,02 | 0,03(**) | 1,03 |
| AP3xPR | -0,65(***) | 0,52 | 1,21(***) | 3,34 | -2,18(***) | 0,11 | -3,70(***) | 0,02 |
| CPxIC | -0,03(**) | 0,97 | 0,01 | 1,01 | 0,00 | 1,00 | -0,01 | 0,99 |
| CPsxCP | 0,50(***) | 1,64 | 0,55(***) | 1,73 | 1,13(***) | 3,09 | 1,18(***) | 3,26 |
| CPxCI | 0,00(*) | 1,00 | 0,00 | 1,00 | 0,00 | 1,00 | 0,00 | 1,00 |
| CPxPR | -0,01 | 0,99 | 0,01 | 1,01 | 0,00 | 1,00 | -0,04(**) | 0,96 |
| CAxCE | -0,61(***) | 0,54 | -0,30(*) | 0,74 | -0,14 | 0,87 | -0,05 | 0,95 |
| CAxCP | -0,46(***) | 0,63 | -0,01 | 0,99 | -0,62(***) | 0,54 | -2,11(***) | 0,12 |
| CAxCI | 0,01(*) | 1,01 | 0,01 | 1,01 | 0,01 | 1,01 | 0,03(***) | 1,03 |
| CAxPR | 0,32(***) | 1,38 | 0,76(***) | 2,15 | 0,47(***) | 1,60 | 0,97(***) | 2,63 |

***p<0,001, **p <0,01, *p <0,05. CE (input centrality), CC (closeness centrality), CI (centrality of intermediation), PR (PageRank), TE (period of work experience), CP (content publication), AP (professional activity: 1-teacher, 2-economist, 3-consultant/investor) and CA (audience communication).

# 7. Conclusions: Future directions

Blogs are as much organizations as organized within networks [11], [17], [19], fails to unite these two arguments to provide a more complete explanation regarding this phenomenon [3]. In order to resolve this limitation, a parallel was established between different perspectives of Network and Organizational theories to develop a theoretical conciliatory model. We found bloggers' characteristics and managerial functions that have explanatory power to attractiveness variation, beyond any environmental "determinism". The effects of connection preference were identified, but contrary to that suggested by [2], these outcomes are not immune to managerial behavior. These network effects are not irreversible [3], [21] on the blogsphere.

Besides proposing a new perspective to understand blog context, it was shown that these approaches are not exclusive but rather complementary. Astley & Van de Ven [7], were already arguing that ecological and contingent approaches can offer better explanations for the interactions between organizations and contexts, an explanation so far missing in the context of digital work and blogs.

In practice, this study helps blog managers or interested companies when it shows that, in spite of contingentialism, there are certain desirable characteristics and managerial functions that bloggers can provide to acquire or maintain attractiveness. Initially, because of its environmental nature, it is essential that a blogger does not just wait passively for readers to appear, but to get actively and strategically involved in attracting a wider audience. In addition, it is vital that a blogger administers the recommendations received from other blogs, balances the number of publications issued and communicates with readers through feedback.

It is also important to investigate in the future, if the blogs with a higher structural capital, or higher attractiveness are influenced by other factors such as the support of media corporations, investors and others. Blogs can be an integrated part of a corporation, a setting interesting for a complementary future study.

As one of our research limitations with regards to data collection, the restrictions imposed by the very manner in which an egocentric network is constructed, should be mentioned. By choosing an alter-blog, the network was automatically linked to its own connections, so that different choices for this actor involved different types of network analysis. In addition, this method implied uncertainty, as initially it was not known how many and which actors would take part in the study. As a result, blogs that did not present public profiles were disregarded due to the impossibility of obtaining the necessary information.

With regards to the external limitation of this research, it should be highlighted that, in spite of structural capital measures typically being used as performance predictors [21], [27], other studies exist (e.g. [5]) that use these as determinants, that is to say, as a measure of the performance itself. There is still the possibility of understanding these measures as an outcome of attractiveness, since it is possible to argue that greater visibility, for example, arises from greater attractiveness. Finally, there exist potential relationships between the variables that were not explored. An audience, for example, can affect a blogger's decisions about what to write, how much to write and how to write. In order to add this type of analysis to our theoretical model, we suggest a future study is carried out, involving research that addresses the involvement of the reader's and the blogger's perspectives, which would demand the use of different types of research instruments and levels of analysis.